\documentclass{elsart}

 \usepackage{graphics}
 \usepackage{graphicx}
 \usepackage{epsfig}
\def\bye{\end{document}}
\usepackage{amssymb}

\begin{document}

\begin{frontmatter}



\title{The QCD critical end point in the SU(3) Nambu--Jona-Lasinio model}


\author[label1,label3]{P. Costa}
\ead{pcosta@teor.fis.uc.pt}
\author[label1]{C. A. de Sousa}
\ead{celia@teor.fis.uc.pt}
\author[label1]{M. C. Ruivo}
\ead{maria@teor.fis.uc.pt}
\author[label2,label3]{Yu. L. Kalinovsky}
\ead{kalinov@nusun.jinr.ru}
\address[label1]{Departamento de F\'{\i}sica, Universidade de Coimbra, P - 3004 - 516 Coimbra, Portugal}
\address[label2]{Universit\'{e} de Li\`{e}ge, D\'{e}partment de Physique B5, Sart Tilman, B-4000, Li\`{e}ge 1, Belgium}
\address[label3]{Laboratory of Information Technologies, Joint Institute for Nuclear Research, Dubna, Russia}

\date{\today}
\begin{abstract}

We study the chiral phase transition at finite temperature $T$ and baryonic chemical potential $\mu_B$ within the framework of the SU(3) Nambu-Jona-Lasinio (NJL) model.
The QCD critical end point (CEP) and the critical line at finite $T$ and $\mu_B$ are investigated: the study of physical quantities, such as the baryon number susceptibility and the specific heat in the vicinity the CEP, will provide relevant information concerning the order of the phase transition. The class of the CEP is determined by calculating the critical exponents of those quantities.
\end{abstract}
%

\end{frontmatter}
%

\section{Introduction}

It is commonly accepted that the vacuum of quantum chromodynamics (QCD) undergoes a phase transition to the quark gluon plasma (QGP) at high temperature and/or quark chemical potential. Such a new state of matter is experimentally studied in on-going heavy ion collisions at CERN, Brookhaven and JIRN \cite{Sissakian:2006}.

The discussion about the existence of a tricritical point (TCP) or a CEP  is also a topic of recent interest. As is well known,  a TCP  separates the first order transition at high chemical potential from the second order transition at high
temperatures. If the  second order transition is replaced by a smooth crossover, a CEP which separates the two lines is found.
The existence of the CEP in QCD was suggested at the end of the eighties \cite{Asakawa:1989NPA,Barducci:1994PRD}, and its properties have been studied since then (for a review see Ref. \cite{Stephanov:2004,Casalbuoni:2006}).

The most recent lattice results with $N_f=2+1$ staggered quarks of physical masses indicate the location of the CEP at $T^{CEP}=162\pm2\mbox{MeV},\,\mu^{CEP}=360\pm 40\mbox{ MeV}$ \cite{Fodor:2004JHEP}, however its exact location is not yet known (it depends strongly of the mass of the strange quark). At the CEP the phase transition is of second order, belonging to the three-dimensional Ising universality class, and this kind of phase transitions are characterized by long-wavelength fluctuations of the order parameter.

The possible signatures of the CEP in heavy-ion collisions have been studied in detail in \cite{Stephanov:2004,Stephanov:1998PRL,Stephanov:1999PRD}.
In heavy-ion collision experiments, fluctuations of $T$ and $\mu_B$, can be found, respectively, in event-by-event fluctuations of $p_T$ spectra or in event-by-event fluctuations in the baryon number to pion ratio \cite{Stephanov:1998PRL,Hatta:2003PRL}.
The event-by-event fluctuations of $T$ can be related to the heat capacity:
if the specific heat $C$ diverges, as the CEP is approached from either the left or the right, the fluctuations of $T$ are suppressed.
On the other hand, the event-by-event fluctuations of $\mu_B$ can be related to the baryon number susceptibility, $\chi_B$, which is also divergent. This implies that fluctuations of $\mu_B$ are also suppressed at the critical point\footnote{The divergence of $\chi_B$ is directly related to an anomaly in the event-by-event fluctuation of baryon number: in a heavy-ion collision experiment it is expected that the event-by-event fluctuation of the proton number is relatively enhanced for collisions which have passed in the vicinity of the CEP or the TCP. It is also expected an increase in event-by-event fluctuations at low $p_T$ near the CEP \cite{Stephanov:2004,Stephanov:1998PRL}.}.

As pointed out in Ref. \cite{Hatta:2003PRD}, the critical region around the CEP is not pointlike but has a richer structure. The critical region is defined as the region where the mean field theory of phase transitions breaks down and nontrivial critical exponents emerge. The size of this critical region is important for future searches for the CEP in heavy ion-collisions \cite{{Nonaka:2005PRC}}.

It is also expected that the strange quark should have an important effect on the position of the TCP and the CEP.
At finite $T$ and zero $\mu_B$, in the limiting case with $m_u=m_d=0$ and infinite strange quark mass $m_s$, the chiral phase transition is likely to be of second order and the static critical behaviour is expected to belong to the universality class of the Heisenberg $O(4)$ model in three dimensions \cite{Pisarski:1984PRD}.
When $m_s$ is finite and less than some critical value $m_{s}^{crit}$, the second order transition becomes of first order. This leads to a tricritical point in the $T-m_s$ plane \cite{Wilczek:1992IJMP}.

Some studies have been done in the SU(2) sector \cite{Hatta:2003PRD,Schaefer:2006} but less attention has been given to the effects of the strange quark \cite{Rajagopal:1999NPA}. In this paper we aim at investigating the chiral phase transition and the CEP in quark matter with strange quarks.


\section{Model calculations}\label{model}

We perform our calculations in the framework of the three--flavor NJL model \cite{njl,klevkuni,Rehberg:1995PRC,Buballa:2004PR}, including the determinantal 't Hooft interaction that breaks the $U_A(1)$ symmetry, which Lagrangian reads
\begin{eqnarray} \label{lagr}
{\mathcal L} &=& \bar{q} \left( i \partial \cdot \gamma - \hat{m} \right) q
+ \frac{g_S}{2} \sum_{a=0}^{8}
\Bigl[ \left( \bar{q} \lambda^a q \right)^2+
\left( \bar{q} (i \gamma_5)\lambda^a q \right)^2
 \Bigr] \nonumber \\
&+& g_D \Bigl[ \mbox{det}\bigl[ \bar{q} (1+\gamma_5) q \bigr]
  +  \mbox{det}\bigl[ \bar{q} (1-\gamma_5) q \bigr]\Bigr] \, .
\end{eqnarray}
Here $q = (u,d,s)$ is the quark field with three flavors, $N_f=3$, and three colors, $N_c=3$. $\hat{m}=\mbox{diag}(m_u,m_d,m_s)$ is the current quark mass matrix and $\lambda^a$ are the Gell--Mann matrices, a = $0,1,\ldots , 8$, ${\lambda^0=\sqrt{\frac{2}{3}} \, {\bf I}}$. The three momentum integrals are regularized by the cutoff $\Lambda$, and a standard set of parameters \cite{klevkuni,Costa:2003PRC} given by $\Lambda =602.3$ MeV, $g_S\Lambda^2 = 3.67$, $g_D \Lambda^5 = -12.36$,  $m_u = m_d = 5.5$ MeV and  $m_s = 140.7$ MeV allow to reproduce  the following vacuum observables: $M_{\pi} = 135.0$ MeV, $M_K   = 497.7$ MeV, $f_\pi = 92.4$ MeV and $M_{\eta'}= 960.8$ MeV. With this set of parameters we also obtain $\left\langle\bar{q}_{u}\,q_u\right\rangle =
\left\langle\bar{q}_{d}\,q_d\right\rangle = - (241.9 \mbox{ MeV})^3$ and $\left\langle\bar{q}_{s}\,q_s\right\rangle = - (257.7 \mbox{MeV})^3$, for the quark condensates and $M_u=M_d=367.7$ MeV and $M_s=549.5$ MeV, for the constituent quark masses.

The fundamental relation is provided by the baryonic thermodynamic potential
\begin{equation}\label{tpot}
    \Omega (\mu_i ,T)= E- TS - \sum_{i=u,d,s} \mu _{i} N_{i}\,,
\end{equation}
from which the relevant equations of state for the entropy $S$, the pressure $P$ and the particle number $N_i$ can be calculated as usually (the expressions are given in Sec. IV of Ref. \cite{Costa:2003PRC}). So we take the temperature $T$, the volume $V$ and the chemical potential of the $i$-quark ($\mu_i$) as the full independent state variables, and a grand canonical approach is applied to our model of strong interacting matter. It can simulate either a region in the interior of a neutron star, or a hot and dense fireball created in a heavy-ion collision. Since electrons and positrons are not involved in the strong interaction, we impose the condition $\mu_e=0$. So we naturally get the chemical equilibrium condition $\mu_u=\mu_d=\mu_s=\mu_B$ that is used along the work.
This condition is only valid around the CEP, where the temperature is expected to be high enough to avoid  color condensation, or at small chemical potential where this phenomena is not yet present.
The baryon number susceptibility is the response of the baryon number density $\rho_B(T, \mu_i)$ to an infinitesimal variation of the quark chemical potential $\mu_i$ \cite{McLerran:1987PRD}:
\begin{equation} \label{chi}
    \chi_B = \frac{1}{3}\sum_{i=u,d,s}\left(\frac{\partial
    \rho_i}{\partial\mu_i}\right)_{T}.
\end{equation}
Here, the quark density is $\rho_i = N_i/V = \frac{ N_c}{\pi^2}\int p^2 dp \left( n_i(\mu_i ,T) -\bar{n}_i(\mu_i ,T)\right)$.
Other relevant observable, in the context of possible signatures for chiral symmetry restoration in the hadron-quark transition and for transition from the hadronic matter to the QGP \cite{McLerran:1987PRD,Asakawa:2000PRL,Blaizot:2001PLB}, is the specific heat which is defined by
\begin {equation}
    C = \frac{T}{V}\left ( \frac{\partial S}{\partial T}\right)_{N_i}
    = \frac{T}{V}\left[\left ( \frac{\partial S}{\partial T} \right)_{\mu_i}
    - \frac{[(\partial N_i/\partial T)_{\mu_i}]^2}{(\partial N_i/\partial \mu_i)_T}\right],
\end {equation}
where we have transformed the derivative $(\partial S/\partial T)_{N_i}$ using the formula of the Jacobian. In fact,  we work in the grand canonical  ensemble where $(T,V,\mu_i)$ are the set of natural independent variables (still holding $N_i$ and $V$ fixed).


\section{The critical end point of QCD}

\begin{figure}[t]
\begin{center}
  \begin{tabular}{cc}
    \hspace*{-0.5cm}\epsfig{file=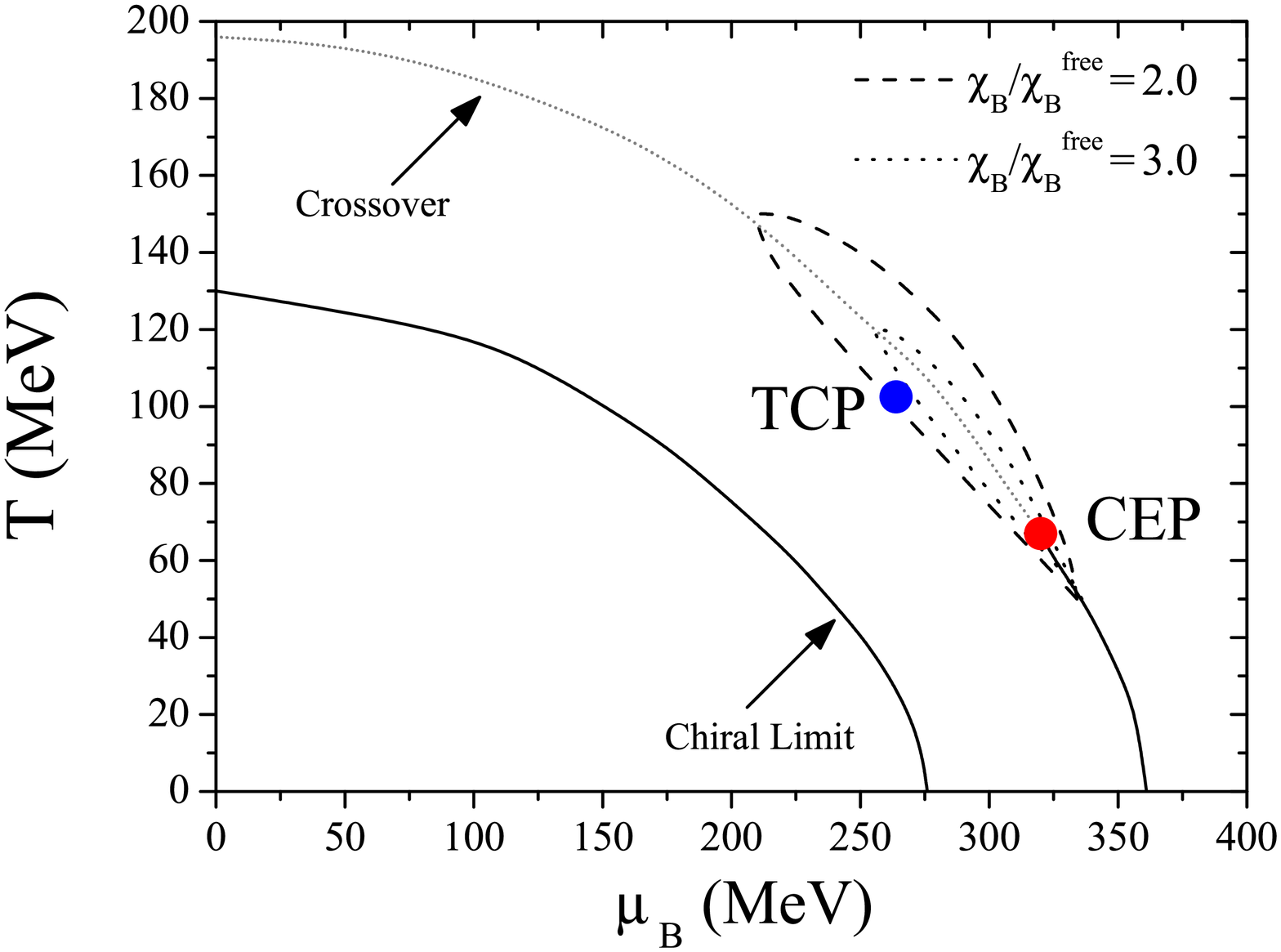,width=7.50cm,height=7cm} &
        \hspace*{-0.75cm}\epsfig{file=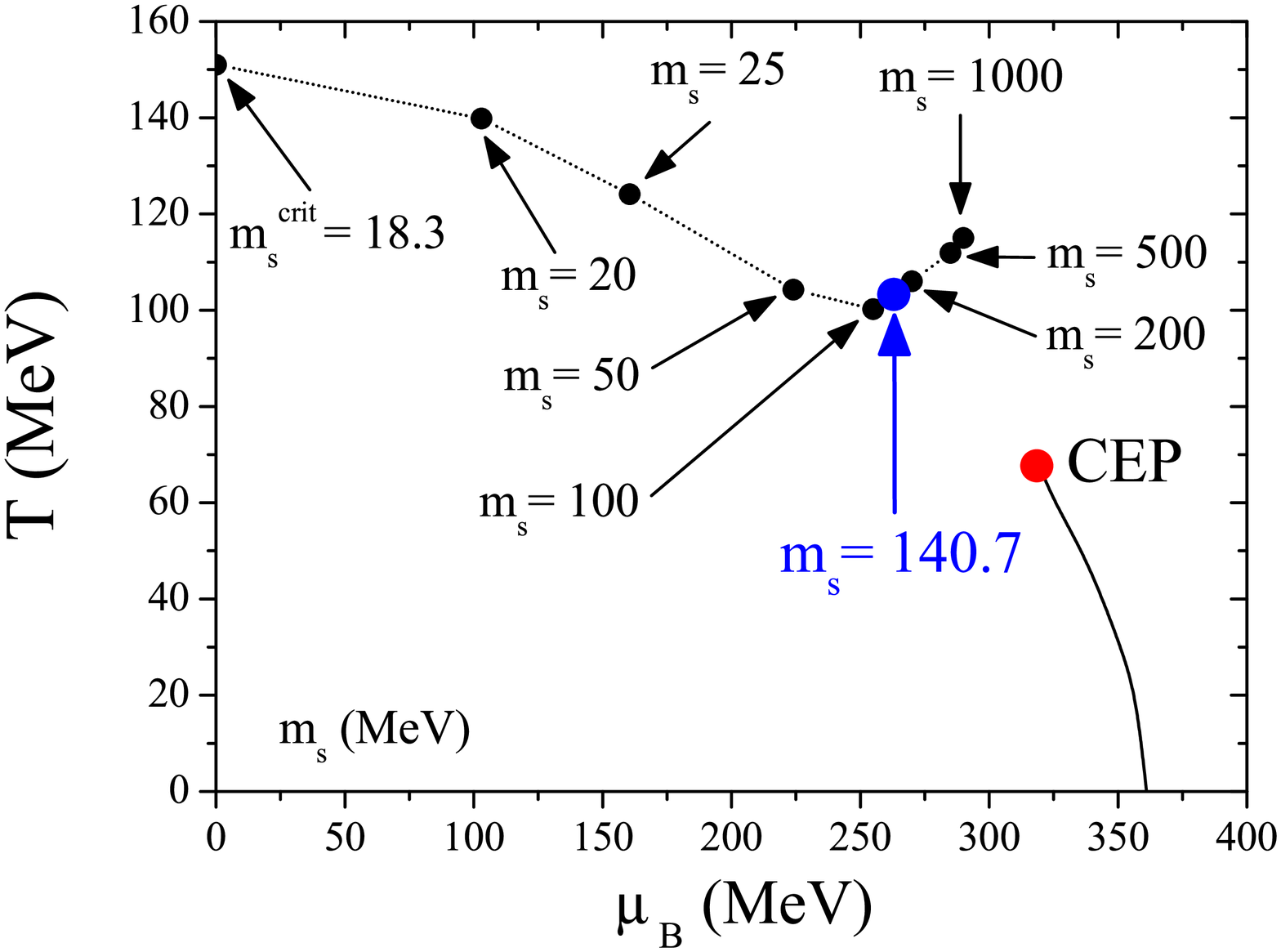,width=7.50cm,height=7cm} \\
   \end{tabular}
\end{center}
\vspace{-0.5cm}
\caption{Left panel: the phase diagram in the SU(3) NJL model. The solid line represents the first order phase transition. The size of the critical region is also plotted for $\chi_B/\chi_B^{free}=2(3)$. Right panel: the phase diagram and the \textquotedblleft line\textquotedblright of TCPs for $m_u=m_d=0$ and different values of $m_s$ (the dotted lines are just drawn to guide the eye).}
\label{Fig:diagfases}
\end{figure}

To study the influence of explicit chiral symmetry breaking on the location of the critical points we vary the current quark masses $m_i$ keeping the other model parameters. The phase diagram for the SU(3) NJL model is presented in Fig. \ref{Fig:diagfases} as a function of $\mu_B$ and $T$, and considering different cases for the current quark masses. The main motivation is the discussion of the critical behavior of the system, starting with the location of the CEP. Using physical values of the quark masses \cite{Rehberg:1995PRC,Costa:2005PRD}: $m_u = m_d = 5.5$ MeV, $m_s = 140.7$ MeV, this point is localized at $T^{CEP}=67.7$ MeV and $\mu_B^{CEP} = 318.4$ MeV ($\rho_B^{CEP}=1.68\rho_0$). We also verified that, contrarily to what happens in the chiral limit for the SU(2) sector where the TCP is found, the NJL model in SU(3), also in the chiral limit ($m_u=m_d=m_s=0$), does not exhibit a TCP \footnote{Both situations are in agreement with what is expected: the chiral phase transition at the chiral limit is of second order for $N_f = 2$ and first order for $N_f\geq3$ \cite{Pisarski:1984PRD}.}: chiral symmetry is restored via a first order transition for all baryonic chemical potentials and temperatures (see left panel of Fig. \ref{Fig:diagfases}).
This pattern of chiral symmetry restoration remains for $m_u=m_d=0$ and $m_s<m_{s}^{crit}$ \cite{Hsu:1998PLB}. In our model we found $m_{s}^{crit}=18.3$ MeV for $m_u=m_d=0$. When $m_s\geq m_{s}^{crit}$, at $\mu_B=0$, the transition is of second order and, as $\mu_B$ increases, the line of the second order phase transition will end in a first order line at the TCP.  Several TCPs are plotted  for different values of $m_s$ in the right panel of Fig. \ref{Fig:diagfases}. As $m_s$ increases, the value of $T$ for this \textquotedblleft line\textquotedblright of TCPs decreases as $\mu_B$ increases getting closer to the CEP and, when $m_{s}=140.7$ MeV, it starts to move away from the CEP. The TCP for $m_{s}=140.7$ MeV is the closest to the CEP and is located at $\mu_B^{TCP}=265.9$ MeV and $T^{TCP}=100.5$ MeV.
If we choose $m_u=m_d\neq0$, instead of second order transition we have a smooth crossover for all the values of $m_s$ and the \textquotedblleft line\textquotedblright of TCPs becomes a \textquotedblleft line\textquotedblright of CEPs.

\section{Behaviour of $\chi_B$ and $C$  in the vicinity of the CEP and their critical exponents}

A bound to the size of the critical region around the CEP can be found by calculating the baryon number susceptibility, the specific heat and their critical behaviours.
If the critical region of the CEP is small, it is expected that most of the fluctuations associated with the CEP will come from the mean field region around the critical region \cite{Hatta:2003PRD}.

\begin{figure}[t]
\begin{center}
  \begin{tabular}{cc}
    \hspace*{-0.5cm}\epsfig{file=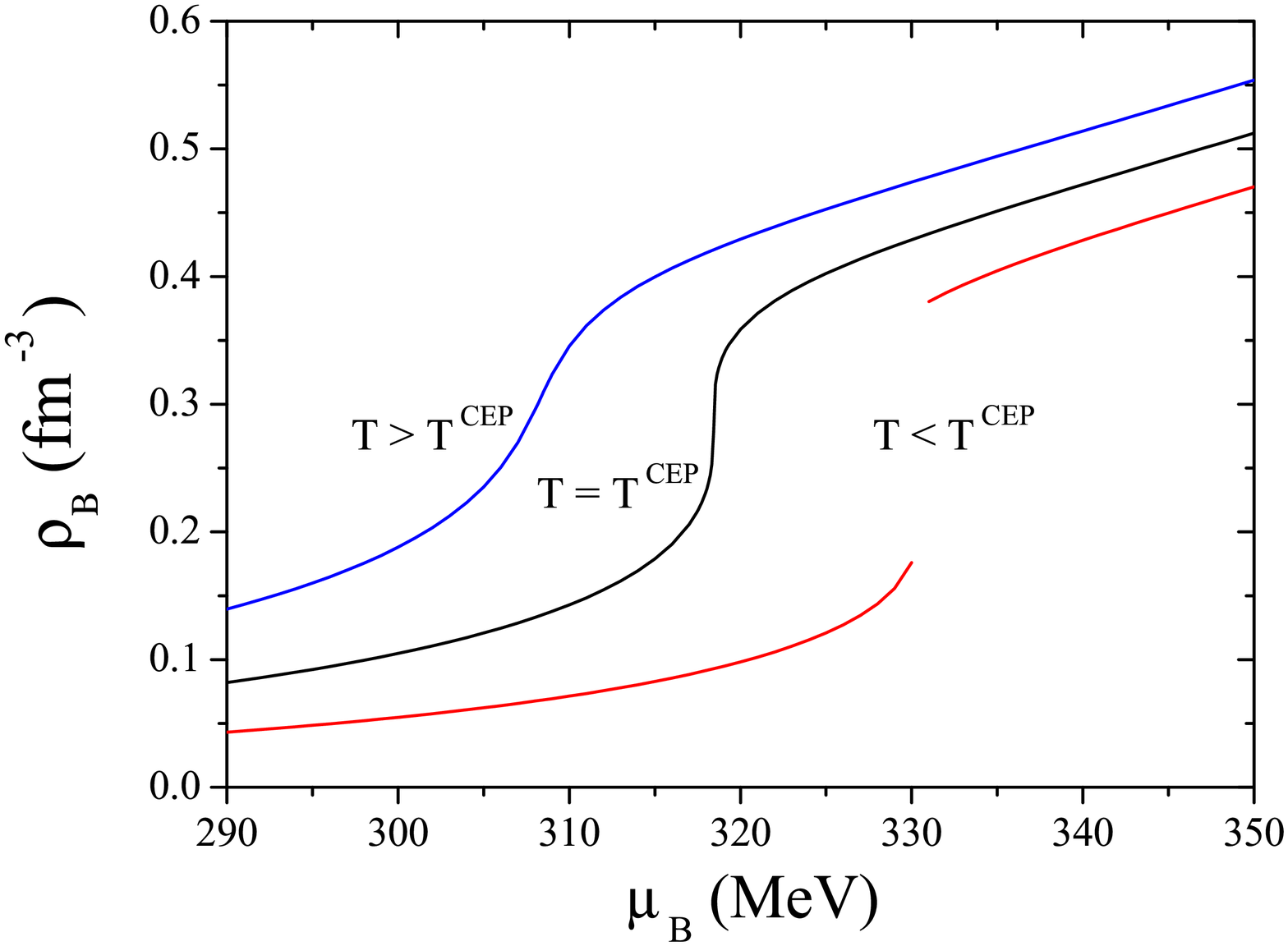,width=7.50cm,height=7cm} &
        \hspace*{-0.75cm}\epsfig{file=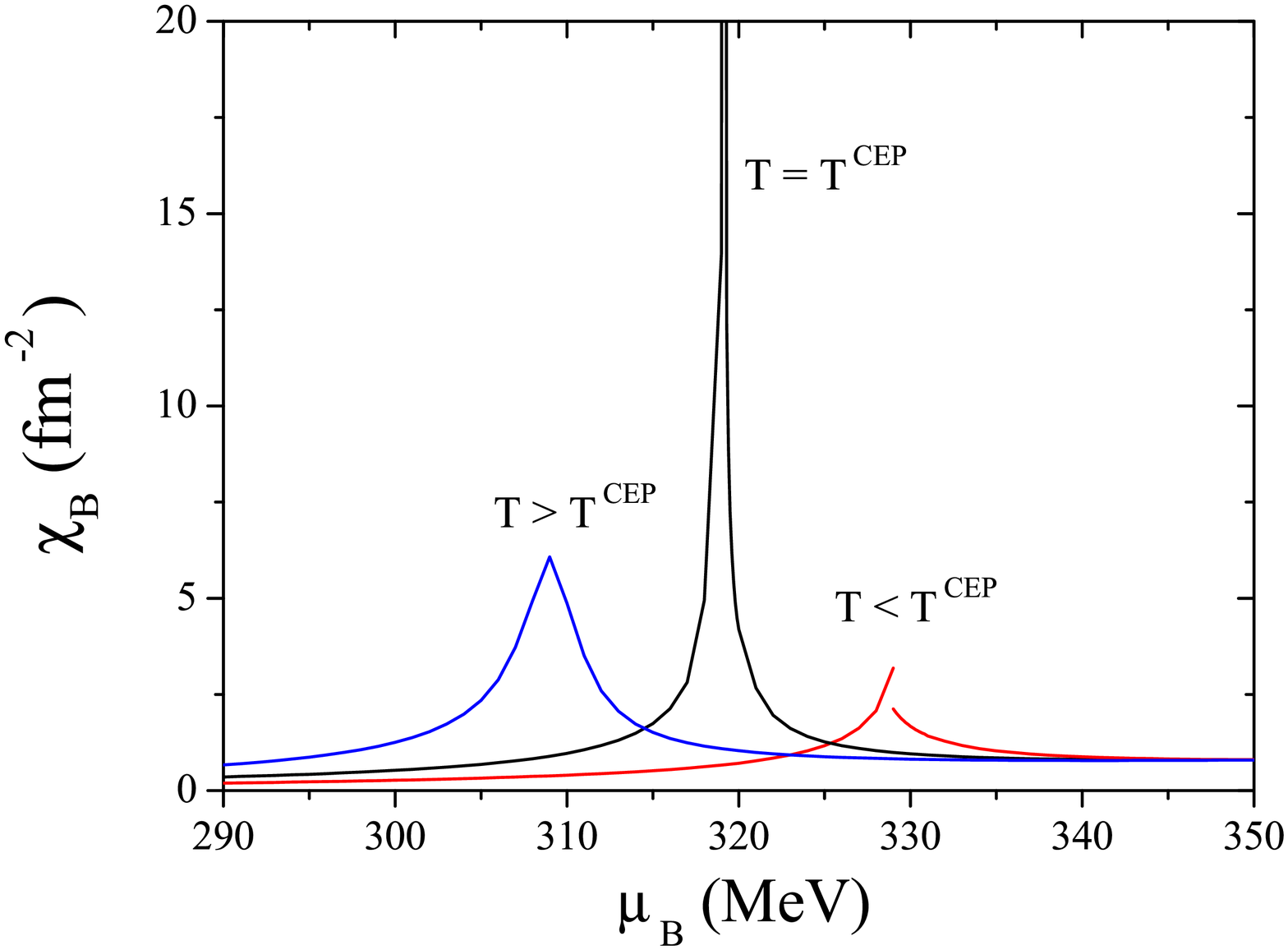,width=7.50cm,height=7cm} \\
   \end{tabular}
\end{center}
\vspace{-0.5cm}
\caption{Baryonic density (left panel) and baryon number susceptibility (right panel) as function of $\mu_B$ for different temperatures around the CEP: $T^{CEP}=67.7$ MeV and $T=T^{CEP}\pm10$ MeV.}
\label{CEPchiT}
\end{figure}

\begin{figure}[tp]
\begin{center}
        \epsfig{file=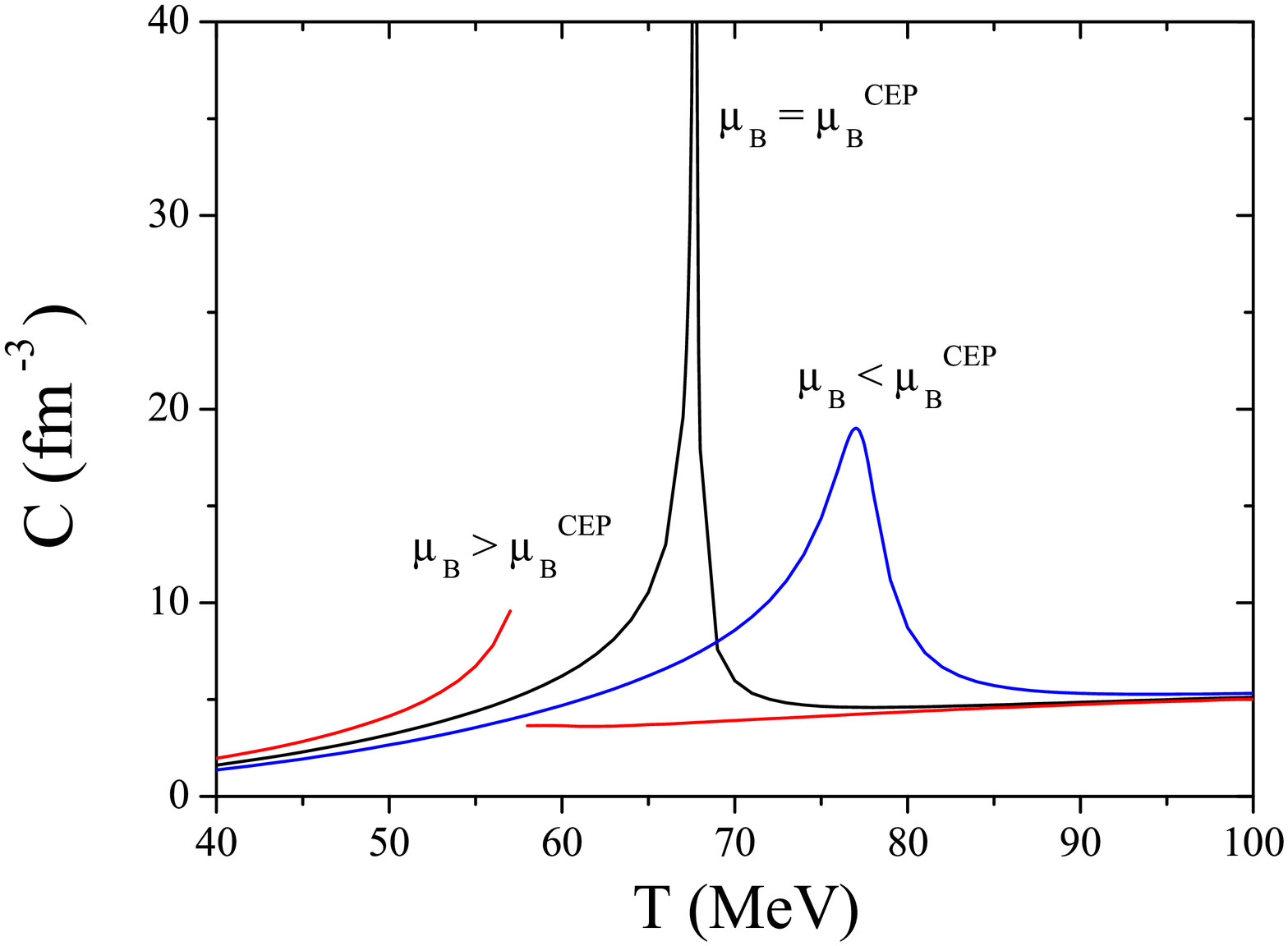,height=7cm,width=7.5cm}
\end{center}
\vspace{-0.5cm}
\caption{Specific heat as a function of $T$ for different values of $\mu_B$ around the CEP ($\mu_B^{CEP}=318.4$ MeV and $\mu_B=\mu_B^{CEP}\pm10$ MeV).}
\label{fig:CEPceT}
\end{figure}

In the left panel of Fig. \ref{CEPchiT} is plotted the baryon number density for three different temperatures around the CEP. For temperatures below $T^{CEP}$ we have a first order phase transition and, consequently, $\chi_B$ has a discontinuity (right panel of Fig. \ref{CEPchiT}). For $T = T^{CEP}$ the slope of the baryon number density tends to infinity at $\mu_B=\mu_B^{CEP}$ which implies a diverging susceptibility (this behaviour was found in \cite{Hatta:2003PRD,Schaefer:2006} using different models in the SU(2) sector).
For temperatures above $T^{CEP}$, in the crossover region, the discontinuity of $\chi_B$ disappears at the transition line, and the density changes gradually in a continuous way as we can see in the right panel of Fig. \ref{CEPchiT}.
%
%
A similar behavior is found for the specific heat for three different chemical potentials around the CEP, as we can observe from Fig. \ref{fig:CEPceT}.

As we have already seen, several thermodynamic quantities diverge at the CEP. In order to make this statement more precise, we will focus on the values of a set of indices, the so-called critical exponents, which describe the behavior near the critical point of various quantities of interest (in our case $\epsilon$ and $\alpha$ are the critical exponents of $\chi_B$ and $C$, respectively). The motivation for this study arises from fundamental phase transition considerations, and thus transcend any particular system. These critical exponents will be determined by finding two directions, temperature-like and magnetic-field-like, in the $T-\mu_B$ plane near the CEP, because, as pointed out in \cite{Griffiths:1970PR}, the form of the divergence depends on the route which is chosen to approach the critical end point.

Starting with the baryon number susceptibility, if the path chosen is asymptotically parallel to the first order transition line, the divergence of $\chi_B$ scales with an exponent $\gamma_B$. In the mean-field approximation it is expected $\gamma_B=1$ for this path. 
For any other path not parallel to the first order line, the divergence scales with the exponent $\epsilon = 1-1/\delta$. Once in the mean-field approximation $\delta=3$, we will have $\epsilon =2/3$ and $\gamma_B > \epsilon$ is verified. The last condition is responsible for the elongation of the critical region, $\chi_B$
being enhanced in the direction parallel to the first order transition line (see Fig. \ref{Fig:diagfases}).
To estimate the critical region around the CEP we can calculate the dimensionless ratio $\chi_B/\chi_B^{free}$ where $\chi_B^{free}$ is obtained taking the chiral limit $m_u=m_d=m_s=0$. Left panel of Fig. \ref{Fig:diagfases} shows a contour plot for two fixed ratios ($\chi_B/\chi_B^{free}=2.0(3.0)$) in the phase diagram around the CEP where we confirm the elongation, in the direction parallel to the first-order transition line, of the region where $\chi_B$ is enhanced.

\begin{figure}[t]
\begin{center}
  \begin{tabular}{cc}
    \hspace*{-0.5cm}\epsfig{file=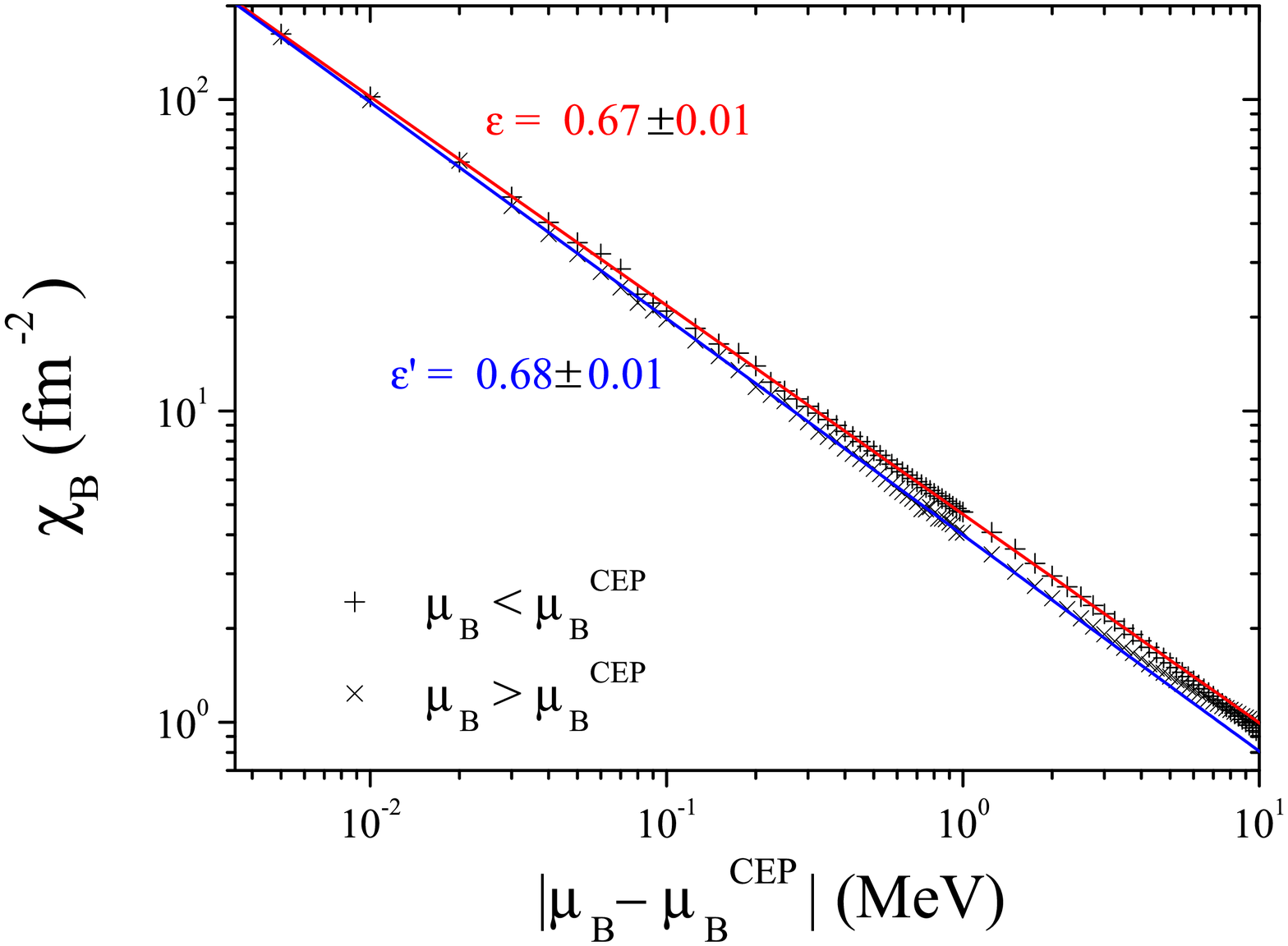,width=7.50cm,height=7cm} &
        \hspace*{-0.75cm}\epsfig{file=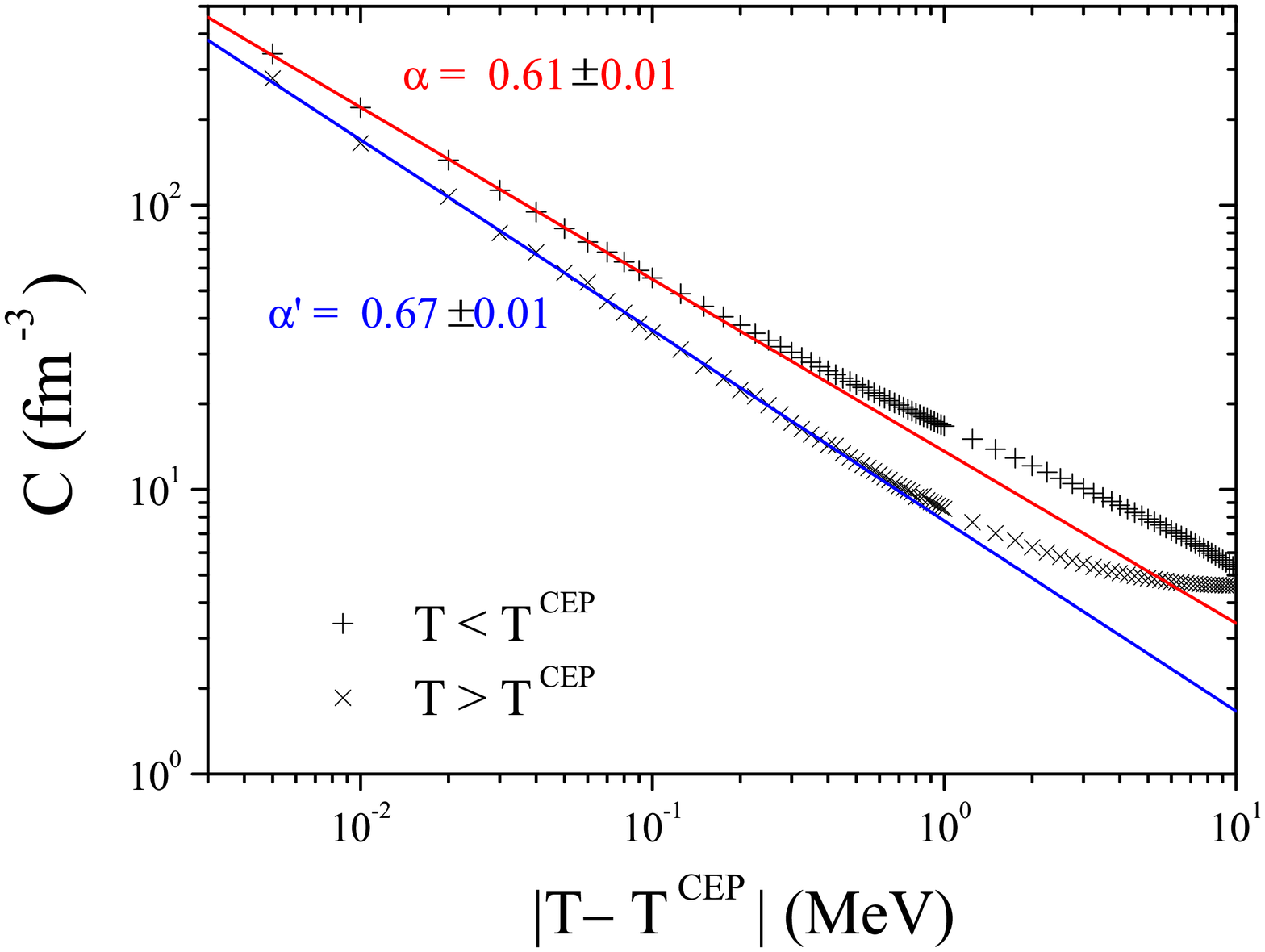,width=7.50cm,height=7cm} \\
   \end{tabular}
\end{center}
\vspace{-0.5cm}
\caption{Left panel: baryon number susceptibility as a function of $|\mu_B-\mu_B^{CEP}|$ at fixed temperature $T^{CEP}=67.7$ MeV. Right panel: specific heat as a function of $|T-T^{CEP}|$ at fixed chemical potential $\mu_B^{CEP}=318.4$ MeV.}
\label{fig:critexp}
\end{figure}

For the baryon number susceptibility we will start with a path parallel to the $\mu_B$-axis in the ($T,\mu_B$)-plane, from lower $\mu_B$ towards the critical $\mu_B^{CEP}=318.4$ MeV, at fixed temperature $T^{CEP}=67.7$ MeV. In Fig. \ref{fig:critexp} we plot $\chi_B$ as a function of $\mu_B$ close to the CEP. Using a linear logarithmic fit
\begin{equation}
	\ln \chi_B = -\epsilon \ln |\mu_B -\mu_B^{CEP} | + c_1 ,
\end{equation}
where the term $c_1$ is independent of $\mu_B$, we obtain $\epsilon = 0.67\pm 0.01$, which is consistent with the mean field theory prediction $\epsilon = 2/3$.

Once there is no reason why the critical exponent should be equal for both regions, below and above $\mu_B^{CEP}$, we also study the baryon number susceptibility from higher $\mu_B$ towards the critical $\mu_B^{CEP}$. The logarithmic fit used now is $\ln \chi_B = -\epsilon' \ln |\mu_B -\mu_B^{CEP}| + c'_1$. Our result shows that $\epsilon' = 0.68\pm 0.01$ which is very near the value of $\epsilon$. This means that the size of the region we observe is approximately the same independently of the direction we choose for the path parallel to the $\mu_B$-axis.

Paying attention to the specific heat around the CEP, we have used a path parallel to the $T$-axis in the ($T,\mu_B$)-plane from lower (higher) $T$ towards the critical $T^{CEP}=67.7$ MeV at fixed $\mu_B^{CEP}=318.4$ MeV. In Fig. \ref{fig:critexp} we plot $C$ as a function of $T$ close to the CEP in a logarithmic scale.
Now, we see that for the region $T<T^{CEP}$ we have a nontrivial critical exponent $\alpha=0.61\pm 0.01$\footnote{We use the linear logarithmic fit $\ln C = -\alpha \ln |T -T^{CEP}| + c_2$ where the term $c_2$ is independent of $T$.}. This value of $\alpha$ is not in agreement with what was suggested by universality arguments in \cite{Hatta:2003PRD}: it is expected that $\chi_B$ and $C$ should be essentially the same near the TCP and the CEP which implies $\alpha=\epsilon=2/3$.
One possible interpretation of this result could be the effect of the hidden TCP on the CEP, as it was already seen in Refs. \cite{Hatta:2003PRD,Schaefer:2006}, and which influence could be stronger in the specific heat rather than in the baryon number susceptibility. However, this explanation is not valid in the framework of the NJL model once, at the TCP, the value of $\alpha$ is already not consistent with the respective mean field value.

To support the last statement, let us analyze the behavior of the specific heat around the TCP when $m_s=140.7$ MeV ($T^{TCP}=100.5$ MeV, $\mu_B^{TCP}=265.9$), the nearest TCP from the CEP (see right panel of Fig. \ref{Fig:diagfases}). Using a path parallel to the $T$-axis from lower $T$ towards the critical $T^{TCP}$, at fixed chemical potential $\mu_B^{TCP}$, we find $\alpha=0.45\pm0.01$. This result, in spite of being close, is not in agreement with the respective mean field value ($\alpha=1/2$). We notice that the inconsistency with the mean field values only occurs for the specific heat. In fact, it is found that the critical exponent for $\chi_B$ ($\gamma_B$ once we are in the TCP) whose value, $\gamma_B=0.50\pm0.02$, is in agreement with the respective mean
field value ($\gamma_B=1/2$).

Nevertheless we observe that the values of $\alpha$ in the TCP and in the CEP are consistent within the NJL model.
We also stress that the universality arguments are so general that they give no quantitative results and, due to the lack of information from the lattice simulations, they should be confronted with model calculations. The eventual difference between the values of the $C$ and $\chi_B$ critical exponents can be interesting in heavy-ion
collisions experiments. Finally, for the region $T>T^{CEP}$ the critical exponent is $\alpha'=0.67\pm 0.01$ which is compatible with the value of $\epsilon'$. This means that the specific heat is sensitive to the way we approach the CEP.


\section{Summary and conclusions}

We have analyzed the phase diagram in the SU(3) NJL which reproduces the essential features of QCD: a first order phase transition for low temperatures and the existence of the CEP. Contrarily to what happens in the SU(2) sector, in the chiral limit ($m_u=m_d=m_s=0$) we do not find a  TCP in the NJL model, which agrees with what  is
expected: the chiral phase transition at $m_i = 0$ is  second order for $N_f = 2$ and first order for $N_f\geq3$. When $m_u=m_d=0$ and $m_s>m_{s}^{crit}$ ($m_{s}^{crit}=18.3$ MeV) the transition is second order ending in a first order line at the TCP. As $m_{s}$ increases we have a \textquotedblleft line\textquotedblright of
TCPs. For $m_u=m_d\neq0$ there is a crossover for all the values of $m_s$ and the \textquotedblleft line\textquotedblright of TCPs becomes a \textquotedblleft line\textquotedblright of CEPs. The location of the CEP depends strongly of the strange quark mass.
Around the CEP we have studied the baryon number susceptibility and the specific heat which are related with event-by-event fluctuations of $\mu_B$ or $T$ in heavy-ion collisions. In the NJL model, for $\chi_B$, we conclude that the obtained critical exponents are consistent with the mean field values $\epsilon=\epsilon'=2/3$ (the NJL model only produces mean field behaviors). From our study of the critical exponent for the specific heat, we conclude that $\alpha$ is different from $\epsilon$. More relevant information about the CEP can be obtained from the spectral functions and the isentropic trajectories in its vicinity. This work is in progress.


\begin{flushleft}
\textbf{Acknowledgments}
\end{flushleft}

One of the authors (P.C.) acknowledges helpful conversations with Yoshitaka Hatta.
Work supported by grant SFRH/BPD/23252/2005 from F.C.T. (P. Costa), by grant RFBR 06-01-00228 (Yu. Kalinovsky), Centro de Física Teórica and by F.C.T. under project POCI/FP/63945/2005.


\end{document}